\documentclass[fleqn,usenatbib]{mnras}

\usepackage{newtxtext,newtxmath}

\usepackage[T1]{fontenc}

\DeclareRobustCommand{\VAN}[3]{#2}
\let\VANthebibliography\thebibliography
\def\thebibliography{\DeclareRobustCommand{\VAN}[3]{##3}\VANthebibliography}

\usepackage{graphicx}	
\usepackage{amsmath}	

\title[quasar variational slope]{The variational slope of quasar light curves is not a distance indicator}

\author[C. J. Burke et al.]{
Colin J. Burke$^{1}$\thanks{E-mail: colinjb2@illinois.edu}
\\
$^{1}$Department of Astronomy, University of Illinois at Urbana-Champaign, 1002 W. Green Street, Urbana, IL 61801, USA\\
}

\date{Accepted XXX. Received YYY; in original form ZZZ}

\pubyear{2023}

\begin{document}
\label{firstpage}
\pagerange{\pageref{firstpage}--\pageref{lastpage}}
\maketitle

\begin{abstract}
When the time difference quotients, or \emph{variational slopes}, of quasar light curves are plotted against their absolute magnitudes, there is a tight positive correlation of $\sim 0.16$ dex in the variational slope direction or $\sim 0.5$ dex in the absolute magnitude direction. This finding resulted in suggestions that a variational slope -- luminosity relation could be used as a distance indicator. However, I show that this relation can be explained almost entirely from self-correlation with luminosity. After properly accounting for the self-correlation component, the relation has a true scatter of $\sim 1.5$ dex in luminosity, consistent with established correlations for quasar variability amplitudes. Given this large scatter, correlation with variational slope or variability amplitude and luminosity is not by itself a suitable distance indicator for quasars.
\end{abstract}




\section{Introduction} \label{sec:intro}

The $4-6\sigma$ discrepancy between measurements of the cosmological constant $H_0$ with early and late time probes is a pressing problem in cosmology today \citep{DiValentino2021}. Quasars are attractive objects to establish as standard candles because they can be easily observed over a wide range of redshifts from $z \sim 0 - 5$ (e.g., \citealt{Shen2011}). Therefore, extending the distance ladder to these redshifts could reveal new physics or elucidate the discrepancy between early and late time measurements of $H_0$. 

Recent claims of a tight empirical correlation between quasar variability---specifically the time difference quotients or \emph{variational slopes} and luminosity---have generated interest in using quasar light curves to obtain an independent measurement of their luminosity distances \citep{Dai2012,Solomon2022}. In this work, I seek to understand the quasar variational slope within the context of established measures of quasar variability (e.g., \citealt{MacLeod2010}), and find that this variational slope -- luminosity relation can be explained almost entirely from self-correlation with luminosity.

\section{Variational slope analysis} \label{sec:slope}

The flux variational slope of quasar light curves can be defined as the difference quotient:
\begin{equation}
\label{eq:Ndraw}
    s_{F,\ {\rm{obs}}} \equiv \frac{dF}{dt} \approx \frac{F(t + \Delta t) - F(t)}{\Delta t},
\end{equation}
where $t$ is the measurement time and $\Delta t$ is the (uniform) grid spacing in the observed frame. Following \cite{Dai2012,Solomon2022}, this can be measured by fitting straight line segments to quasar light curves within specified windows of length $\sim \Delta t$. The mean variational slope can then be determined from the distribution of variational slopes for each quasar \citep{Solomon2022} or by fitting a line to a new light curve constructed from the straight line fits in each window with each segment shifted to ($t = 0$, $F = 0$) \citep{Dai2012}.  Alternatively, one could compute the gradients of a time-varying model fit to the light curve directly. \cite{Solomon2022} define the rest-frame as $s_F = (1 + z)\ s_{F,\ {\rm{obs}}}$, where $s_{F,\ {\rm{obs}}}$ is measured from the observed-frame luminosity light curve.


\subsection{Proof of self-correlation}
\label{sec:selfcorrelation}

\begin{figure*}
\centering
\includegraphics[width=0.49\textwidth]{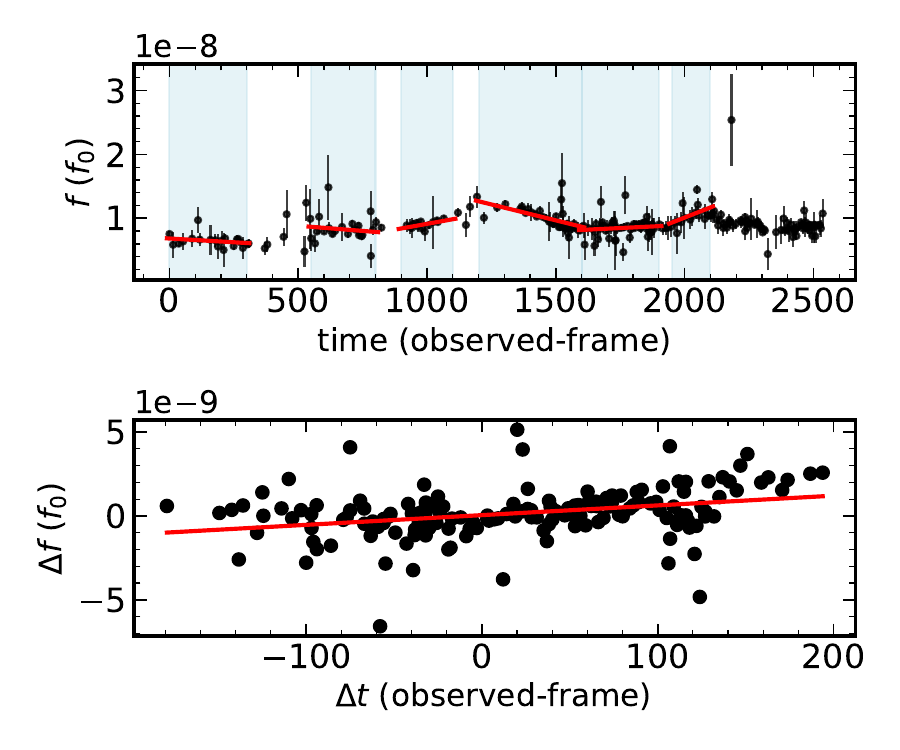}
\includegraphics[width=0.49\textwidth]{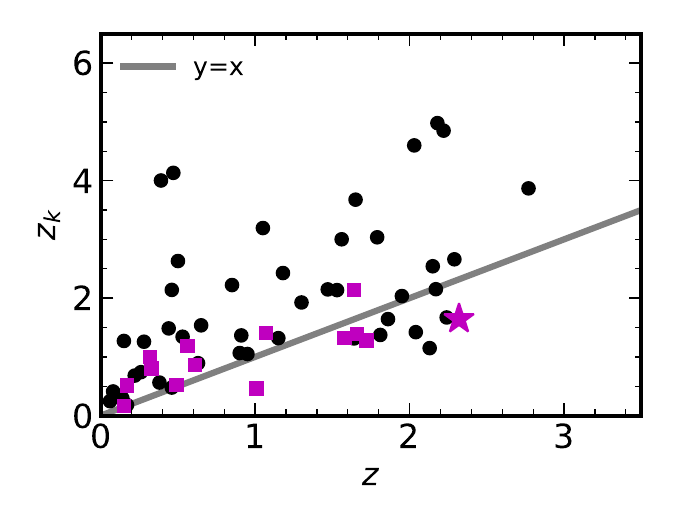}
\caption{\emph{Left:} example variational slope extraction procedure following \citet{Dai2012}. The upper panel shows the light curve for the quasar MACHO 64.8092.454 (black points) and straight line fits to the light curve data (red lines) within specified time windows (light blue). The bottom panel shows a fit to a new light curve constructed from the straight line fits in each window with each segment shifted to ($t = 0$, $F = 0$). \emph{Right:} Resulting predicted redshift $z_k$ vs. true redshift $z$ using the variational slope measurements with Equation~\ref{eq:z}. The magenta squares are the same 13 $V \sim 19$ quasars used in the \citet{Dai2012} analysis. The black circles show the results for the other MACHO quasars. The magenta star is the quasar MACHO 9.5484.258 used as an example by \citet{Dai2012}. \label{fig:slope}}
\end{figure*}

The flux of an astronomical source is related to its magnitude as,
\begin{equation}
\label{eq:flux2mag}
    F = F_0\ 10^{-0.4\ M},
\end{equation}
where $F$ is the luminosity (absolute flux), $M$ is the absolute magnitude, and $F_0$ is the band-specific constant which I set to unity ($F_0 \equiv 1$) for consistency with \cite{Solomon2022}. For small changes in flux, the relation is,
\begin{equation}
\label{eq:dflux2dmag}
    \frac{\Delta F}{F} \approx \frac{\ln 10}{2.5} \Delta M,
\end{equation}
with an analogous relationship between apparent magnitude $m$ and flux $f$.

Quasar rms variability amplitude has a modest dependence on luminosity \citep{MacLeod2010,Suberlak2021}. The empirical relation of \cite{Suberlak2021} is,
\begin{equation}
\label{eq:dm}
    \lvert \Delta m \lvert \propto \left\lvert \frac{\Delta f}{\langle f \rangle} \right\lvert \propto \left\lvert \frac{\Delta F}{\langle F \rangle} \right\lvert \propto 10^{(0.118\pm0.003) \langle M \rangle},
\end{equation}
after ignoring the logarithmic dependence on mass and the small wavelength dependence. Here $m$ is apparent magnitude, $f$ is the apparent flux, $F$ is the absolute flux, and $\langle f \rangle$, $\langle F \rangle$ denote the time-averaged (mean) quantities. The variational slope is then,
\begin{equation}
\label{eq:Ndraw}
    \lvert s_F \lvert \sim \left\lvert \frac{\Delta F}{\Delta t} \right\lvert \propto 10^{(0.118\pm0.003) \langle M \rangle} \langle F \rangle, 
\end{equation}
using Equation~\ref{eq:dm}. Hence, the variational slope is expected to be proportional to absolute flux due to self-correlation. Taking the logarithm and making use of Equation~\ref{eq:flux2mag}, it follows that,
\begin{equation}
\label{eq:logsF}
    \log |s_F| \propto (-0.4 + 0.118\pm0.003) \langle M \rangle,
\end{equation}
for the variational slope. The predicted slope from self-correlation of $\log |s_F| \propto \langle -0.4\ M \rangle$ plus the dependence from \cite{MacLeod2010} is $-0.282 \pm 0.003$, which is in excellent agreement ($< 1\sigma$) with the slope of $-0.2929 \pm 0.0098$ found by \cite{Solomon2022}. It also follows that the scatter of $0.5$ dex in the luminosity direction in a $\log |s_F|$ vs. $\langle M \rangle$ scatter plot is under-estimated by $1$ dex due to self-correlation. In contrast, when the variational slope is computed in magnitudes instead of flux, $s_M$, the self-correlation is proportional to  $\log\ \langle F \rangle$, reducing strength of the self-correlation and increasing the scatter to $\sim1.5$ dex with the self-correlating $\langle F \rangle$ term removed.


\subsection{A universal slope of quasar variability?}

\cite{Dai2012} carried out a similar analysis and found a universal slope of quasar flux variability $k$ instead of a luminosity dependence. The sample of \cite{Dai2012} only includes 13 MACHO quasars \citep{Geha2003} with very well sampled light curves with roughly the same apparent magnitudes of $V\sim19$. I re-did the variational slope analysis following \cite{Dai2012} but for all MACHO quasars using apparent fluxes (i.e., computing $s_f$) with a time window of $\Delta t \sim 90$ days in the rest frame for consistency with \cite{Dai2012}. Taking the average of our apparent flux variational slopes for the same 13 MACHO quasars, I obtain $k = 8.3 \pm 0.7 \times 10^{-11} F_0$ days$^{-1}$, roughly consistent with the \cite{Dai2012} measurement. Clearly, how one measures the variational slope matters, but the systematic differences in measuring the variational slopes should wash out as long as the same method is used. Using Equation~3 of \cite{Dai2012},
\begin{equation}
\label{eq:z}
    z_k = \frac{k}{s_f} -1,
\end{equation}
I show the predicted redshifts for each MACHO quasar in Figure~\ref{fig:slope}. 

In Figure~\ref{fig:slope}, I can roughly recover the true redshifts of the subset of quasars with similar apparent fluxes. However, the predicted redshifts $z_k$ for the full sample of MACHO quasars with a broader range in apparent fluxes does not in general tightly track with the true redshifts using Equation~3 of \cite{Dai2012}. This can be understood as follows. From previous work, $\lvert \Delta f/\langle f \rangle \lvert \propto \lvert \Delta m \lvert$ is approximately constant, with some scatter and only a mild dependence on luminosity (e.g. \citealt{MacLeod2010,Suberlak2021}). Hence, $k \sim |\Delta f/\Delta t| $ is only approximately constant when averaged over a sample of quasars with similar $\langle f \rangle$ values. Hence, when the full sample of quasars is used with a wide range of apparent magnitudes/fluxes, the predicted redshifts are much more uncertain.


Furthermore, the ``rest-frame'' flux variational slope measurements in \cite{Dai2012} are defined as $(1 + z)\ s_{f,\ {\rm{obs}}}$. In addition to the time-dilation factor of $(1 + z)$, a quasar at redshift $z$ has its fluxes scaled by its luminosity distance $d_L$ as $F = f\ 4 \pi d_L^2$. It follows that the rest-frame variational slope should be $s_F = 4 \pi d_L^2\ (1 + z)\ s_{f,\ {\rm{obs}}} = (1 + z)\ s_{F,\ {\rm{obs}}}$. Therefore, even if the intrinsic rest-frame luminosity variations are universal for quasars, i.e., a universal $s_F$, there can be no universal slope $k$ in $(1 + z)\ s_{f,\ {\rm{obs}}}$.

\section{Discussion \& Conclusions}

After reproducing the main results of \cite{Solomon2022} and \cite{Dai2012} and scrutinizing their conclusions, I demonstrate that the observed quasar variational slope is not very predictive of redshift. Using simple arguments, I show that the correlation between variational slope and luminosity found by \cite{Solomon2022} is mostly an artifact from self-correlation, and that it is consistent with established correlations of quasar variability from \cite{MacLeod2010}. It is reasonable to posit that the variations in luminosity may be universal for quasars at fixed luminosity, if it is set by the accretion rate of the quasar \citep{Dexter2011,Meusinger2013}. However, the poorly-understood and large scatter of $\sim 1.5$ dex in the luminosity direction means quasar variability amplitude alone is probably not a suitable cosmological distance indicator. 

\section*{Acknowledgements}

I thank Yue Shen for helpful discussions, and the anonymous referee for comments that improved this work.

\section*{Data Availability}

The MACHO quasar light curves can be found at \url{http://www.astro.yale.edu/mgeha/MACHO}.



\bibliographystyle{mnras}
\bibliography{example} 





\bsp	
\label{lastpage}
\end{document}